\DocumentMetadata{}
\documentclass[sigplan,screen]{acmart}

\usepackage{xspace, xcolor}
\usepackage{enumitem}
\usepackage[noend]{algpseudocode}
\usepackage{algorithm}
\usepackage[caption=false]{subfig}
\usepackage{balance}
\usepackage[all]{nowidow}
\usepackage{listings}
\usepackage{needspace}

\usepackage{tabularx, multirow}
\newcolumntype{L}{>{\raggedright\arraybackslash}X}
\newcolumntype{C}{>{\centering\arraybackslash}X}
\newcolumntype{R}{>{\raggedleft\arraybackslash}X}

\usepackage[]{hyperref}

\newcommand{\NAME}{Cheddar\xspace}

\newcommand{\evk}{$\mathsf{evk}$\xspace}
\newcommand{\evks}{$\mathsf{evk}$s\xspace}
\newcommand{\hrot}{$\mathtt{HRot}$\xspace}
\newcommand{\hconj}{$\mathtt{HConj}$\xspace}
\newcommand{\hadd}{$\mathtt{HAdd}$\xspace}
\newcommand{\pmult}{$\mathtt{PMult}$\xspace}
\newcommand{\hmult}{$\mathtt{HMult}$\xspace}
\newcommand{\boot}{$\mathtt{bts}$\xspace}
\newcommand{\bitprime}[1]{Pr$\sim$#1\xspace}
\newcommand{\efflevel}{$L_\text{eff}$\xspace}
\newcommand{\parf}[1]{\textsc{Par.#1}}
\newcommand{\seqf}[1]{\textsc{Seq.#1}}

\newcommand{\centerdash}{\multicolumn{1}{c}{-}}
\newcommand{\centerdashtwo}{\multicolumn{1}{c}{\multirow{2}{*}{-}}}

\newcommand{\rev}[1]{{\color{black}#1}}

\setlist{leftmargin=*}
\definecolor{codegreen}{rgb}{0,0.6,0}
\definecolor{codepurple}{rgb}{0.58,0,0.82}
\definecolor{keywordblack}{rgb}{0, 0, 0}

\lstdefinestyle{mystyle}{
    commentstyle=\color{codegreen},
    keywordstyle=\color{keywordblack},
    stringstyle=\color{codepurple},
    basicstyle=\linespread{0.9}\scriptsize\ttfamily,     
    breaklines=true,
    keepspaces=true,
    showspaces=false,
    showstringspaces=false,
    showtabs=false,
    tabsize=2,
    language=sh,
    alsoletter={-}
}

\copyrightyear{2026}
\acmYear{2026}
\setcopyright{cc}
\setcctype{by}
\acmConference[ASPLOS '26]{Proceedings of the 31st ACM International Conference on Architectural Support for Programming Languages and Operating Systems, Volume 1}{March 22--26, 2026}{Pittsburgh, PA, USA}
\acmBooktitle{Proceedings of the 31st ACM International Conference on Architectural Support for Programming Languages and Operating Systems, Volume 1 (ASPLOS '26), March 22--26, 2026, Pittsburgh, PA, USA}
\acmDOI{10.1145/3760250.3762223}
\acmISBN{979-8-4007-2165-6/2026/03}

\title{Cheddar: A Swift Fully Homomorphic Encryption Library Designed for GPU Architectures}

\author{Wonseok Choi}
\orcid{0009-0004-0941-4805}
\authornote{Wonseok Choi and Jongmin Kim contributed equally to this work.}
\affiliation{%
 \institution{Seoul National University}
 \city{Seoul}
 \country{Republic of Korea}
}
\email{wonseok.choi@snu.ac.kr}

\author{Jongmin Kim}
\orcid{0000-0003-2937-3073}
\authornotemark[1]
\affiliation{%
 \institution{Seoul National University}
 \city{Seoul}
 \country{Republic of Korea}
}
\email{jongmin.kim@snu.ac.kr}

\author{{Jung Ho} Ahn}
\orcid{0000-0003-1733-1394}
\affiliation{%
 \institution{Seoul National University}
 \city{Seoul}
 \country{Republic of Korea}
}
\email{gajh@snu.ac.kr}

\begin{document}

\begin{abstract}
Fully homomorphic encryption (FHE) frees cloud computing from privacy concerns by enabling secure computation on encrypted data.
However, its substantial computational and memory overhead results in significantly slower performance compared to unencrypted processing.
To mitigate this overhead, we present \NAME, a high-performance FHE library for GPUs, achieving substantial speedups over previous GPU implementations.
We systematically enable 32-bit FHE execution, leveraging the 32-bit integer datapath within GPUs.
We optimize GPU kernels using efficient low-level primitives and algorithms tailored to specific GPU architectures.
Further, we alleviate the memory bandwidth burden by adjusting common FHE operational sequences and extensively applying kernel fusion.
\NAME delivers performance improvements of 2.18--4.45$\times$ for representative FHE workloads compared to state-of-the-art GPU implementations.
\end{abstract}

\ccsdesc[500]{Security and privacy~Cryptography}
\ccsdesc[500]{Computer systems organization~Single instruction, multiple data}

\keywords{Fully homomorphic encryption, Cryptography, GPU, Security}

\renewcommand{\shortauthors}{Wonseok Choi, Jongmin Kim, and Jung Ho Ahn}
\renewcommand{\shorttitle}{Cheddar: A Swift Fully Homomorphic Encryption
Library Designed for GPU Architectures}

\maketitle 

\section{Introduction}
\label{sec:introduction}

The ubiquitous application of cloud services across increasingly diverse domains has led to a rapid expansion in user data collection, raising serious concerns about security and privacy.
Under such circumstances, homomorphic encryption (HE~\cite{2021-standard}) is gaining attention due to its ability to provide strong cryptographic guarantees for user privacy.
HE enables the processing of encrypted data (ciphertexts), allowing clients to offload computations on private data to cloud servers without ever exposing the underlying information.

Fully homomorphic encryption (FHE~\cite{stoc-2009-gentry-fhe}) is a special subclass of HE that supports a mechanism called bootstrapping (\boot).
Without \boot, only a fixed number of operations can be performed on a ciphertext; \boot refreshes the ciphertext to enable practically an infinite number of operations on it.

Despite these strengths, hurdles remain for the real-world use of FHE, as processing encrypted data with FHE is several orders of magnitude slower than unencrypted computation~\cite{access-2021-demystify}.
Fortunately, FHE operations exhibit high degrees of parallelism~\cite{isca-2022-bts}, allowing for significant performance improvements through hardware acceleration using parallel computing.
Previous studies have demonstrated remarkable performance improvements on various hardware systems, including CPUs~\cite{access-2021-demystify, wahc-2021-hexl, discc-2024-fhe-cnn}, GPUs~\cite{hpca-2023-tensorfhe, discc-2024-fhe-cnn, micro-2023-gme, tches-2021-100x, hpca-2025-warpdrive}, FPGAs~\cite{hpca-2019-roy, asplos-2020-heax, hpca-2023-fab, hpca-2023-tensorfhe, hpca-2025-effact}, and custom ASICs~\cite{micro-2021-f1, isca-2022-bts, isca-2022-craterlake, micro-2022-ark, isca-2023-sharp}.

This study focuses on GPU acceleration of CKKS~\cite{asia-2017-ckks}, a prominent FHE scheme that exhibits the highest throughput among FHE schemes~\cite{iacr-2023-demystify-boot}.
CKKS is adequate for various practical applications (e.g., machine learning~\cite{ccs-2024-neujeans, icml-2022-resnet, access-2022-resnet20, aaai-2019-helr, privatenlp-2020-rnn, icml-2023-hetal}) due to its capability to handle real (or complex) numbers with high arithmetic precision.
Prior GPU acceleration studies have shown fruitful results.
While Lee et al.~\cite{icml-2022-resnet} reported that encrypted convolutional neural network (CNN) inference with the ResNet-20 model (CIFAR-10) took 2271 seconds on a single-core CPU, recent work~\cite{hpca-2025-warpdrive} reduces the time to 5.88 seconds ($386\times$ faster) with an NVIDIA A100 GPU.

Despite such impressive results, we identify opportunities for further improvement; adopting \emph{32-bit residue number system (RNS)} is especially promising for GPU architectures, which natively utilize 32-bit datapath for integer operations.
To meet cryptographic requirements, popular FHE libraries~\cite{wahc-2020-lattigo, heaan-latest, wahc-2022-openfhe, iacr-2020-helib, fcds-2017-seal} use 64-bit primes for RNS decomposition
(64-bit RNS).
On GPUs, this would incur a quadratic growth in computational overhead as GPUs emulate 64-bit integer operations with their 32-bit datapath.
Recent studies introduced advanced mechanisms for 32-bit RNS (e.g., rational rescaling~\cite{asplos-2024-bitpacker, ccs-2025-grafting}, see \S\ref{sec:32bit}), which enable equivalent cryptographic parameter choices with only 32-bit primes.

However, previous 32-bit RNS constructions are not ideal for the use with GPUs.
We identify that they either require excessive memory capacity requirements for additional public keys~\cite{asplos-2024-bitpacker} or reduce the parallelizability due to control divergence~\cite{ccs-2025-grafting}.

To overcome these limitations, we introduce \textbf{\NAME}, a swift GPU library for FHE CKKS, featuring a novel systemized 32-bit RNS construction (25-30 prime system).
We develop a method to minimize the number of unique primes used for RNS, which allows using a single highly compatible public key, without control divergences.
We also propose an inverted-terminal data layout (\S\ref{sec:32bit:layout}), which \rev{augments our 32-bit RNS construction by simplifying} contiguous data allocation through appropriate prime ordering.

\rev{\NAME also includes numerous optimizations at various levels.}
We develop optimized low-level kernels for major polynomial operations supporting our 32-bit RNS construction (\S\ref{sec:kernel}).
We implement the kernels with signed Montgomery reduction~\cite{iacr-2018-signed-montgomery} to reduce the number of integer operations and devise optimization methods for number-theoretic transform (NTT) and basis conversion (BConv), which can be selectively applied according to the computational and memory characteristics of a specific GPU device. 

While our optimized 32-bit kernels effectively reduce the cost of compute-intensive operations, memory-bound automorphism and element-wise operations remain as performance bottlenecks.
To address this, we apply extensive kernel fusion techniques, categorized into sequential and parallel fusion (\S\ref{sec:seq}).
For sequential fusion, we propose the kernel fusion methods that completely eliminate the cost of operations.
For parallel fusion, we deeply analyze FHE operational sequences across various scopes to identify common parallel execution patterns.
\rev{Although our work is not the first to apply kernel fusion to FHE~\cite{tches-2021-100x, hpca-2025-warpdrive}, we fully draw on its capability by extensively reordering and splitting the operational sequences, maximizing} the fusion opportunities.

\NAME is a full-fledged FHE library enabling the translation of high-level programmer codes into fully optimized GPU programs.
In addition to the aforementioned optimizations, \NAME features various high-level FHE mechanism implementations, GPU kernel fine-tuning capability, and state-of-the-art FHE algorithms~\cite{acns-2022-sparseboot, eurocrypt-2021-efficient, rsa-2020-better, micro-2023-mad, micro-2022-ark}.

While we have introduced various distinguished features of \NAME, the core strength of \NAME can be summarized rather briefly: \textbf{it is simply fast.}
We develop highly efficient implementations of practical FHE applications that demonstrate significantly faster execution times than state-of-the-art GPU implementations~\cite{tches-2021-100x, discc-2024-fhe-cnn, hpca-2023-tensorfhe, heaan-latest, hpca-2025-warpdrive, micro-2023-gme} and even custom FPGA designs~\cite{hpca-2023-fab, hpca-2023-poseidon, hpca-2025-effact}.
For instance, \NAME reduces the latency of complex applications like encrypted CNN inference to a practical level on a single GPU.

The main contributions of this work are as follows: 
\begin{itemize}[nosep]
\item We propose the 25-30 prime system, a novel 32-bit RNS design with an inverted-terminal data layout, to enable systematic and efficient FHE execution on GPUs.

\item We develop highly optimized 32-bit GPU kernels using signed Montgomery reduction and architecture-aware optimizations to enhance computational efficiency and memory usage in core FHE operations.

\item We apply kernel fusion extensively, reordering and splitting operational sequences to mitigate memory bandwidth bottleneck.

\item \NAME achieves speedups of 2.18--4.45$\times$ over a state-of-the-art GPU library, reducing encrypted ResNet-20 inference latency to 0.72 seconds on a single RTX 5090 GPU.
\end{itemize}
\section{Background}
\label{sec:background}

We describe FHE with an emphasis on its computational aspects.
Table~\ref{tab:notation} summarizes key notations and symbols.

\subsection{CKKS Fully Homomorphic Encryption (FHE)}
\label{sec:background:ckks}

FHE supports direct computation on encrypted ciphertexts through cryptographic mechanisms.
CKKS~\cite{asia-2017-ckks} supports the encryption of a message $\mathbf{u} \in \mathbb{C}^{N/2}$, a complex vector of length $N/2$.
$\mathbf{u}$ is first encoded as $\langle\Delta\cdot\mathbf{u}\rangle \in \mathcal{R}_Q$, a polynomial in an integer ring $\mathcal{R}_Q = \mathbb{Z}_Q[X]/(X^N + 1)$ with modulus $Q$ and degree $N$.
\textbf{Scale} $\Delta$ is multiplied to $\mathbf{u}$ during encoding to cope with errors.
Higher $\Delta$ leads to higher precision of the outcome \cite{asia-2017-ckks}.
Finally, $\langle\Delta\cdot\mathbf{u}\rangle$ is encrypted into a ciphertext $[\!\langle \Delta \cdot \mathbf{u} \rangle\!]$ by using a random $a$, a secret $s$, and an error $e$:
\begin{equation}
\label{eq:encryption}
[\!\langle \Delta \cdot \mathbf{u} \rangle\!] = (b, a) = (-a \cdot s + \langle \Delta \cdot \mathbf{u} \rangle + e, a) \in \mathcal{R}_Q^2
\end{equation}

A few basic mechanisms of CKKS are introduced. $\odot$ represents element-wise multiplication (mult) and $\ll$ represents cyclic rotation of a vector.
An evaluation key (\evk) consists of $\mathtt{dnum}$ pairs of polynomials in the larger ring $\mathcal{R}_{PQ}$~\cite{rsa-2020-better}; i.e., $\mathsf{evk} \in \mathcal{R}_{PQ}^{2 \times \mathtt{dnum}}$.

\begin{itemize}
\item $\mathtt{HAdd}([\!\langle \Delta \cdot \mathbf{u} \rangle\!], [\!\langle \Delta \cdot  \mathbf{v} \rangle\!]) = [\!\langle \Delta \cdot (\mathbf{u} + \mathbf{v}) \rangle\!]$.
\item $\mathtt{PAdd}([\!\langle  \Delta \cdot \mathbf{u} \rangle\!], \langle  \Delta \cdot \mathbf{v} \rangle) = [\!\langle  \Delta \cdot (\mathbf{u} + \mathbf{v}) \rangle\!]$.
\item $\mathtt{HMult}([\!\langle  \Delta \cdot \mathbf{u} \rangle\!], [\!\langle  \Delta \cdot \mathbf{v} \rangle\!], \mathsf{evk}_{\odot}) = [\!\langle  \Delta^2 \cdot (\mathbf{u} \odot \mathbf{v}) \rangle\!]$.
\item $\mathtt{PMult}([\!\langle \Delta \cdot \mathbf{u} \rangle\!], \langle \Delta \cdot \mathbf{v} \rangle) = [\!\langle \Delta^2 \cdot (\mathbf{u} \odot \mathbf{v}) \rangle\!]$. 
\item $\mathtt{HRot}([\!\langle \Delta \cdot \mathbf{u} \rangle\!], r, \mathsf{evk}_{r}) = [\!\langle \Delta \cdot (\mathbf{u} \ll r) \rangle\!]$.
\end{itemize}
\hmult and \hrot need \evks for computation, where a separate $\mathsf{evk}_r$ is required for every unique \hrot distance $r$.

\subsubsection*{Residue number system (RNS)}

To avoid costly big-integer computations required for large modulus ($Q \sim 2^{1200}$)~\cite{eurocrypt-2018-heaanboot} in FHE, RNS is used to handle the large $Q$ efficiently~\cite{sac-2018-frns-ckks}.
$Q$ is decomposed into a product of smaller primes that fit within the word size (e.g., 32 or 64 bits):
$Q = \prod_{i=0}^{L-1} q_i$,
allowing each polynomial in $\mathcal{R}_Q$ to be represented as $L$ residue polynomials:
\begin{equation*}
a = (a[0], \cdots, a[L - 1]), \text{ where } a[i] = a \text{ mod } q_i.
\end{equation*}
Then polynomial operations are performed limb-wise; e.g.,
\begin{equation*}
a \cdot b = (a[0] \cdot b[0] \text{ mod } q_0, \cdots, a[L - 1] \cdot b[L - 1] \text{ mod } q_{L - 1}).
\end{equation*}
With RNS, a polynomial can be regarded as an $L \times N$ matrix of small integers, each row of which corresponds to a limb.
$P$ (used for \evks) is handled similarly by setting $P = \prod_{i=0}^{\alpha - 1} p_i$.

\setlength{\tabcolsep}{4pt}
\begin{table}
\caption{Notations and symbols.}
\vspace{-0.05in}
\label{tab:notation}
\begin{tabularx}{0.99\columnwidth}{lX}
\toprule
Notation & Explanation\\
\midrule
$N$ & Polynomial ring degree (typically $2^{16}$).\\
$Q$ | $q_i$ | $\tau_i$ & Modulus | main (RNS) primes | terminal (RNS) primes. $Q=\prod_i\tau_i\cdot\prod_iq_i$ for each level (\S\ref{sec:32bit}).\\
$P$ | $p_i$ | $\alpha$ & Auxiliary modulus | auxiliary (RNS) primes | the number of $p_i$'s (fixed). $P=\prod_{i=0}^{\alpha-1}p_i$.\\
$\mathcal{R}_Q$ & Cyclotomic polynomial ring $\mathbb{Z}_Q[X]/(X^N + 1)$.\\
$\langle\Delta\cdot\mathbf{u}\rangle$ & Encoding of $\mathbf{u}\in\mathbb{C}^{N/2}$ (not encrypted).\\
$\Delta$ & Scale multiplied to $\mathbf{u}$ during encoding (e.g., $2^{40}$).\\
$[\!\langle\Delta\cdot\mathbf{u}\rangle\!]$ & Ciphertext ($\in \mathcal{R}_{Q}^{2}$) encrypting $\langle\Delta\cdot\mathbf{u}\rangle$.\\
\evk & Evaluation key ($\in \mathcal{R}_{PQ}^{2\times\mathtt{dnum}}$ for $\mathtt{dnum}$~\cite{rsa-2020-better}).\\
$L_\text{eff}$ & Effective level. The level after bootstrapping.\\
\bottomrule
\end{tabularx}
\end{table}
\setlength{\tabcolsep}{6pt}

\subsubsection*{Rescaling}

Multiplicative mechanisms (\hmult and \pmult) produce ciphertexts with squared scales ($\Delta^2$).
We need to reduce the scale back to $\Delta$ for further computation.
For that purpose, rescaling, an approximate division by $\Delta$, is performed~\cite{asia-2017-ckks}.
For $[\!\langle \Delta^2\cdot\mathbf{u} \rangle\!] \in \mathcal{R}_Q^2$, rescaling reduces the modulus from $Q$ to the next modulus $Q/\Delta$ and produces $[\!\langle \Delta\cdot\mathbf{u} \rangle\!] \in \mathcal{R}_{Q/\Delta}^2$.
For every multiplicative mechanism, this process is repeated with the modulus $Q/\Delta$, $Q/\Delta^2$, and so on.

RNS requires $Q/\Delta$ to be also constructed with RNS primes.
However, as the RNS primes are discrete, the modulus cannot exactly be reduced to $Q/\Delta$ but to $Q'\simeq Q/\Delta$, which results in the scale to be $\Delta'=\Delta^2/(Q/Q')$~\cite{rsa-2022-reckks}.
We aim at maintaining $\Delta'\simeq\Delta$, which requires a careful RNS construction (\S\ref{sec:32bit}).

\subsubsection*{Level and bootstrapping (\boot)}

We associate each modulus with an index called level (e.g., $Q_3$ for 3 levels), which is equal to the number of rescaling applicable to the current modulus of a ciphertext.
At level 0, we have dissipated all the modulus and cannot perform rescaling anymore.
\boot is performed on ciphertexts at level 0 to restore the modulus~\cite{eurocrypt-2018-heaanboot}.
\boot is a complex mechanism including dozens of \hmult and \hrot evaluations.
The frequency of \boot is determined by the level after \boot, which is referred to as \textbf{effective level} (\efflevel)~\cite{isca-2023-sharp}.
Maximizing \efflevel is crucial for FHE performance.

\subsection{Ring Polynomial Operations}
\label{sec:background:ring}

Only the following four types of polynomial operations are sufficient to comprise CKKS mechanisms~\cite{hpca-2025-anaheim}.

\subsubsection*{Basis conversion (BConv)} 

BConv is performed to match the modulus between polynomials before computation.
For example, we can generate $\hat{a} \in \mathcal{R}_P$ from $a \in \mathcal{R}_Q$ by performing the following for each $p_i$ ($i = 0, 1, \dots, \alpha - 1$):
\begin{equation} \label{eq:bconv}
\hat{a}[i] = \textstyle\sum_{j=0}^{L - 1} ( a[j] \cdot (Q/q_j)^{-1} \text{ mod } q_j ) \cdot (Q/q_j) \text{ mod } p_i.
\end{equation}

\subsubsection*{Number-theoretic transform (NTT)}
Polynomial mult can be expensive as it is equivalent to performing a (negacyclic) convolution between length-$N$ vectors of coefficients, where $N$ is typically as large as $2^{16}$.
To alleviate its cost, NTT, a Fourier transform variant for $\mathcal{R}_Q$, is utilized.
With RNS, NTT can be applied to each limb separately if the $q_i$'s satisfy the following condition:
\begin{equation}
\label{eq:ntt-prime}
    q_i \equiv 1 \text{ (mod } 2N).
\end{equation}
After NTT is applied to each limb, polynomial mult boils down to element-wise mult between $L \times N$ matrices.
Well-known fast Fourier transform (FFT) techniques are used to perform NTT with $\mathcal{O}(N\log N)$ complexity for each limb.

\subsubsection*{Automorphism ($\phi_r$)}
Automorphism is an operation found in \hrot.
For automorphism, we shuffle the columns of a polynomial according to the following one-to-one mapping, which moves the i-th column to the j-th position:
\begin{equation}
\label{eq:automorphism}
j = \phi_r(i) = ( ( (2i + 1) \cdot 5^{-r} \text{ mod } 2N ) - 1 ) / 2.
\end{equation}
We extend the use of the symbol $\phi_r$ to also represent automorphism on a polynomial $a$ as $\phi_r(a)$.

\subsubsection*{Element-wise operations}

The rest of the operations are all performed in an element-wise manner; e.g., element-wise addition and mult between $L\times N$ matrices (polynomials).

\subsection{ModSwitch Routine}
\label{sec:background:modsw}

\begin{algorithm}[t]
\small
\caption{Computation of ModSwitch}
\label{alg:modsw}
\begin{algorithmic}[1]

\Require $a[\alpha][N] \in \mathcal{R}_Q$
\Ensure $\hat{a}[L][N] \in \mathcal{R}_P$ 
\State Initialize $temp[L][N]$ and $\hat{a}[\alpha][N]$ with zero values
\State $temp \gets \text{INTT}(a)$
\For{$j \gets 0$ \textbf{to} $\alpha - 1$} \Comment{{\footnotesize BConv const mult}}
    \State $temp[j] \gets temp[j] \cdot (Q/q_j)^{-1} \text{ mod } q_j$
\EndFor
\For{$i \gets 0$ \textbf{to} $\alpha - 1$} \Comment{{\footnotesize BConv matrix mult}}
\For{$j \gets 0$ \textbf{to} $L - 1$}
    \State $\hat{a}[i] \gets \hat{a}[i] + temp[j] \cdot (Q/q_j) \text{ mod } p_i$
\EndFor
\EndFor
\State $\hat{a} \gets \text{NTT}(\hat{a})$
\end{algorithmic}
\end{algorithm}

Hereafter, we regard that a polynomial is in the NTT-applied form by default.
However, prior to performing BConv on a polynomial, it needs to be brought back to the original form by performing inverse NTT (INTT).
This creates a common computational routine of INTT $\rightarrow$ BConv $\rightarrow$ NTT, which is called ModSwitch~\cite{pldi-2020-eva}.
ModSwitch is the most compute-intensive component of CKKS, encompassing nearly all NTT and BConv operations~\cite{hpca-2025-anaheim}.

An exemplar ModSwitch procedure from $\mathcal{R}_Q$ to $\mathcal{R}_P$ is shown in Alg.~\ref{alg:modsw} (ModSwitch can be performed for various input and output moduli pairs, not limited to $Q$ and $P$).
After INTT, BConv is performed in two steps: a per-limb scalar multiplication step (BConv const mult) and a matrix multiplication with precomputed constants (BConv matrix mult).
Finally, NTT is performed.

The mechanisms in \S\ref{sec:background:ckks} consist of complex combinations of ModSwitch routine and other operations in \S\ref{sec:background:ring}.
For example, $\mathtt{HRot}([\!\langle \Delta \cdot \mathbf{u} \rangle\!], r, \mathsf{evk}_{r})$ follows the sequence: ModUp $\to$ KeyMult $\to$ MAC $\to$ automorphism $\to$ ModDown.
Here, ModUp and ModDown are variants of ModSwitch that change the modulus from Q to PQ and vice versa, respectively. KeyMult is an element-wise operation between the polynomial $a \in \mathcal{R}_{PQ}$ and $\mathsf{evk}_r$.
MAC is constant-polynomial mult and add, which is also element-wise operation.

\subsection{GPU Execution Model}
\label{sec:background:gpu}

GPUs are highly parallel computing systems composed of many streaming multiprocessors (SMs), each with numerous int32, fp32, fp64, and tensor cores~\cite{cse-2022-h100}. Programmers write a parallel function, called a kernel, which is executed by many threads on a GPU.
Threads are grouped into thread blocks and further into warps (32 threads), which are scheduled by warp schedulers across the SM sub-partitions.

When a warp is stalled by a memory or data dependency, the scheduler can switch to another ready warp to keep the cores busy. To maintain high utilization, thousands of warps must be launched per kernel. This high degree of thread-level parallelism aligns well with FHE workloads, where independent operations over RNS limbs can be executed in parallel.

\section{32-bit Residue Number System (RNS)}
\label{sec:32bit}

\begin{figure}[t]
    \centering
    \includegraphics[width=0.9\linewidth]{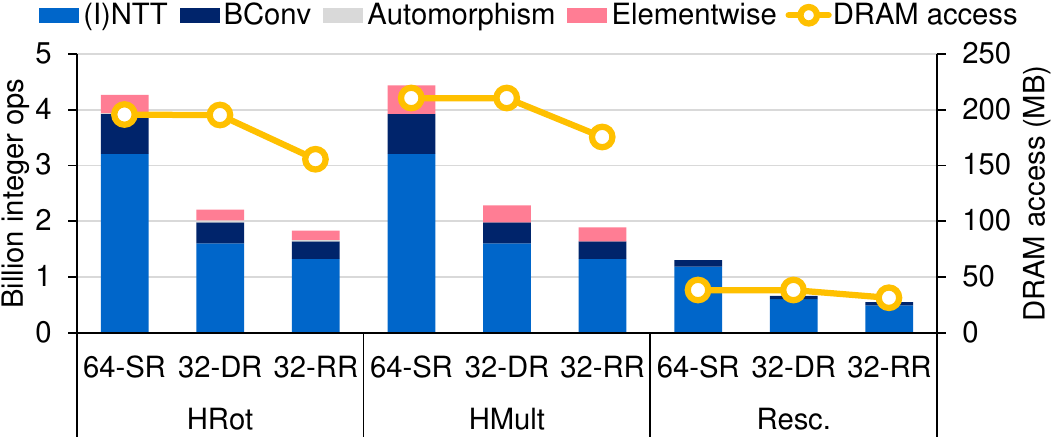}
    \Description{}
    \caption{Integer operation counts and DRAM access amounts of \hrot, \hmult, and rescaling on an RTX 4090 GPU. We used 11-level parameter sets with different word sizes (64/32) and rescaling methods (SR/DR/RR). $N=2^{16}$. $\alpha=6$ (64-bit word) or $\alpha=12$ (32-bit word) was used.}
    \vspace{-0.05in}
    \label{fig:32bit-vs-64bit}
\end{figure}

\begin{figure*}[t]
    \centering
    \includegraphics[width=0.99\linewidth]{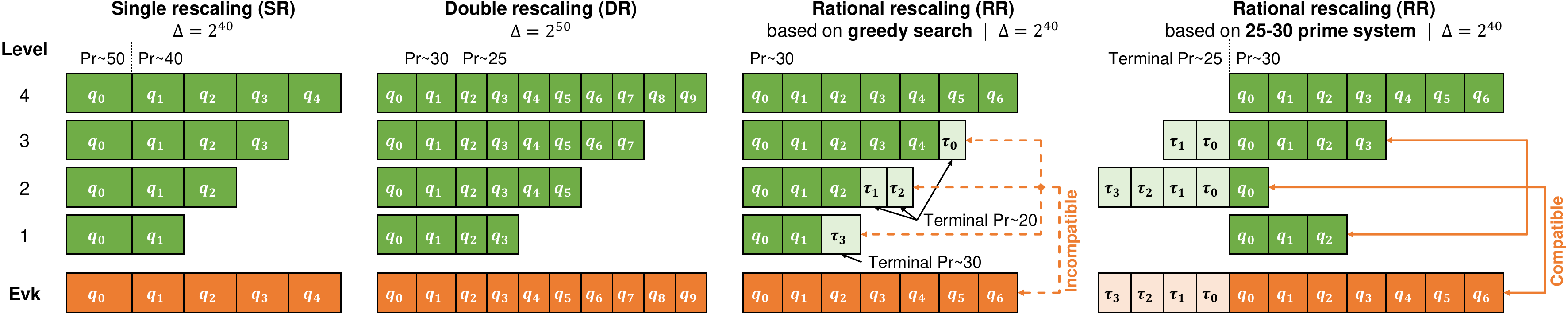}
    \vspace{-0.03in}
    \Description{}
    \caption{RNS constructions (choice of RNS primes comprising $Q$) for non-\boot levels 1--4 and \evk ($P$-part omitted) based on various rescaling methods. \bitprime{k} represents primes sufficiently close to $2^k$.}
    \label{fig:rescaling}
    \vspace{-0.03in}
\end{figure*}

As GPUs utilize int32 cores for integer operations, we choose RNS primes smaller than $2^{31}$, which is the maximum value enabling easy modular addition with 32-bit integer arithmetic.
This 32-bit RNS setting requires up to 2$\times$ (linear) more number of limbs to represent a polynomial compared to conventional 64-bit RNS settings used in popular FHE libraries~\cite{wahc-2020-lattigo, heaan-latest, wahc-2022-openfhe, iacr-2020-helib, fcds-2017-seal}.
However, integer operations with larger word sizes require emulation with int32 operations, whose complexity increases quadratically.
Overall, as shown in Fig.~\ref{fig:32bit-vs-64bit} (64-SR vs. 32-DR), 32-bit RNS requires significantly fewer integer operations for the same mechanism.
Also, the linear increase in the number of limbs does not inflate the amount of DRAM access and rather enhances the degree of parallelism, which leads to higher performance due to a more balanced job distribution on GPUs.

\subsection{Prior RNS Construction Methods}

There are several ways to implement 32-bit RNS based on different rescaling methods.
Using an optimized 32-bit RNS construction with a proper rescaling method can further enhance FHE performance by reducing both computation and DRAM access amounts (32-DR vs. 32-RR in Fig.~\ref{fig:32bit-vs-64bit}).
In this section, we use an example of maintaining $\Delta \simeq 2^{40}$, a typical scale value, throughout a sequence of rescaling.
We refer to the primes sufficiently close to $2^k$ as \bitprime{$k$}.

\subsubsection*{Single and double rescaling}

To maintain $\Delta \simeq 2^{40}$, the easiest solution is to select the RNS primes comprising $Q$ ($q_i$'s except $q_0$) as \bitprime{40}.
Then, we can discard one prime for every rescaling, performing an approximate division by $q_i\simeq\Delta$ for each level, which allows us to maintain the scale close to our target $2^{40}$.
This is called single rescaling (SR).

We may also make the RNS primes satisfy $q_{2i} \cdot q_{2i + 1} \simeq 2^{40}$ ($i > 0$) and discard two primes for every rescaling~\cite{isca-2023-sharp, wahc-2023-32bit}, which we call double rescaling (DR).
However, double rescaling is not scalable for many levels due to the scarcity of primes that satisfy Eq.~\ref{eq:ntt-prime}.
We observe that only one prime smaller than $2^{20}$ satisfies Eq.~\ref{eq:ntt-prime} for $N=2^{16}$, which means double rescaling can be performed only once for $\Delta\simeq2^{40}$.
Thus, double rescaling is only viable for large scales (e.g., $\Delta\simeq2^{50}$ in Fig.~\ref{fig:rescaling}).
Both single and double rescaling are not adequate for 32-bit RNS construction with $\Delta\simeq2^{40}$.

\subsubsection*{Rational rescaling}

BitPacker~\cite{asplos-2024-bitpacker} addresses this problem by adding new RNS primes during rescaling, which is generally referred to as rational rescaling (RR)~\cite{ccs-2025-grafting}.
An exemplar 32-bit RNS construction with BitPacker is shown in Fig.~\ref{fig:rescaling} (RR based on greedy search).
First, all the RNS primes are selected as \bitprime{30}.
For the first rescaling, two \bitprime{30} ($q_5$, $q_6$) are discarded and one \bitprime{20} ($\tau_0$) is added, resulting in the scale being adjusted by roughly $2^{-2\cdot30+20}=2^{-40}$.
For the next rescaling, two \bitprime{30} ($q_3$, $q_4$) as well as $\tau_0$ are discarded and two new \bitprime{20} ($\tau_1$, $\tau_2$) are added ($2^{-2\cdot30-20+2\cdot20}=2^{-40}$).
BitPacker greedily searches the most adequate $\tau_i$'s, which are referred to as \textbf{terminal primes}, for each level to maintain $\Delta\simeq2^{40}$.
RR also allows compact storage of polynomials by composing the modulus with the smallest possible number of RNS primes.
From now on, $q_i$'s denote the main \bitprime{30} primes and $Q = \prod_i \tau_i \cdot \prod_i q_i$ for each level.

\subsection{25-30 Prime System}
\label{sec:32bit:prime-system}

\subsubsection*{Evaluation key (\evk) compatibility problem}

BitPacker keeps adding new primes found by greedy search, which becomes problematic for \evk preparation.
\evks are typically prepared for the top level.
When using a top-level \evk at lower levels, we may truncate the \evk without any computation to adjust its modulus.
Fig.~\ref{fig:rescaling} shows that single or double rescaling allows such truncation.
However, BitPacker cannot do the same due to the terminal primes, requiring additional measures, such as preparing \evks for every possible level.
In the typical parameter set, the key size of $\mathsf{evk}_{\odot}$ with BitPacker blows up by approximately five times compared to that of a single top-level key~\cite{ccs-2025-grafting}.

Grafting~\cite{ccs-2025-grafting} addresses this problem by using a power-of-two RNS base (e.g., $2^{30}$), which is highly amenable to the truncation, instead of the terminal primes.
However, Grafting requires special care of the power-of-two RNS base, which is not adequate for the parallel execution model of GPUs.
For example, (I)NTT with an RNS base $2^{30}$ requires embedding the limb into $\mathcal{R}_{q^*}$ for a large ${q^*}>N\cdot2^{60}$, necessitating costly 128-bit execution on GPUs.

\setlength{\tabcolsep}{2pt}
\begin{table}[t]
\caption{An exemplar 32-bit RNS construction with our 25-30 prime system, where levels 0--4 are non-\boot levels ($L_\text{eff}=4$), and the upper levels are reserved for \boot.}
\vspace{-0.07in}
\label{tab:prime-system}
\begin{tabularx}{0.99\columnwidth}{c|CCCCC|CCCCC}
\toprule
 Level & 0 & 1 & 2 & 3 & 4 & 5 & 6 & 7 & 8 & $\cdots$\\
 \midrule
 \# of \bitprime{25} ($\tau_i$) & 2 & 0 & 4 & 2 & 0 & 1 & 2 & 3 & 4& $\cdots$\\
 \# of \bitprime{30} ($q_i$) & 0 & 3 & 1 & 4 & 7 & 8 & 9 & 10 & 11 & $\cdots$\\
 \midrule
 $\log_2 Q$ & 50 & 90 & 130 & 170 & 210 & 265 & 320 & 375 & 430 & $\cdots$\\
 $\log_2 \Delta$ & - & 40 & 40 & 40 & 40 & 55 & 55 & 55 & 55 & $\cdots$\\
 \bottomrule
\end{tabularx}
\vspace{-0.08in}
\end{table}
\setlength{\tabcolsep}{6pt}

\subsubsection*{25-30 prime system}

To overcome the limitations, we introduce 25-30 prime system, our novel 32-bit RNS construction with a systemized rational rescaling approach.
Instead of selecting the terminal primes separately for each level, we choose them from a fixed list of \bitprime{25} ($\tau_i$'s).
We find a ``rescaling cycle'' to minimize the number of \bitprime{25}; an exemplar rescaling cycle is shown in Table~\ref{tab:prime-system} and Fig.~\ref{fig:rescaling}.
For $\Delta\simeq2^{40}$, we construct a cycle of 1) discarding three \bitprime{30} while adding two \bitprime{25} ($2^{-3\cdot30+2\cdot25}=2^{-40}$), 2) performing 1) once more at the next level, and 3) discarding four \bitprime{25} and adding two \bitprime{30} ($2^{-4\cdot25+2\cdot30}=2^{-40}$) at the level after next.
We can easily find such cycles for typical power-of-$2^5$ scales ($2^{30}$--$2^{50}$) with no more than five \bitprime{25} used for each case.

The primes used at each level are selected in a fixed order from carefully chosen lists of \bitprime{30} and \bitprime{25}, which limits the divergence of scales~\cite{rsa-2022-reckks} to less than 0.1 bits; the scale stays in between $2^{39.9}$ and $2^{40.1}$ in our $\Delta\simeq2^{40}$ example.
Although our 25-30 prime system does not support an arbitrary scale, the provided choices are fine-grained enough to support the precision required by various data types; otherwise, we can also construct similar prime systems (e.g., 24-30).

We set the modulus for \evks as $PQ_\text{max}$, where $Q_\text{max}$ is the product of all the \bitprime{30} and \bitprime{25} used (e.g., $Q_\text{max}=\prod_{i=0}^3\tau_i\cdot\prod_{i=0}^6q_i$ in Fig.~\ref{fig:rescaling}), which makes each \evk compatible with all levels.
As the rescaling cycle limits the number of terminal primes used, the gap between $Q_\text{max}$ and the top-level $Q$ value is small.
Still, unlike single or double rescaling, we are wasting the modulus budget; security constraints limit $PQ_\text{max}$ to certain values depending on $N$ (e.g., $PQ_\text{max}\le2^{881}$ for $N=2^{15}$)~\cite{2021-standard}.
Therefore, although small, we want to get rid of the gap to allow ciphertexts to use as large modulus as possible, which leads to more available levels.

\subsubsection*{Double rescaling for \boot levels}

We remove the modulus gap by selecting primes differently for the upper levels reserved for \boot (\boot levels).
In FHE CKKS, the highest levels are only used during \boot.
Larger scales, typically around $2^{50}$--$2^{60}$~\cite{eurocrypt-2021-efficient}, are used at \boot levels.
We sample the primes for \boot levels based on double rescaling, which is adequate for such large scales.
While doing so, we ensure that all the RNS primes are eventually included at the final top level.
For example, if we are performing \boot at levels 5--19 in addition to the construction in Fig.~\ref{fig:rescaling}, we may force levels 5--8 to use double rescaling with primes $\{\tau_0, q_7\}$,  $\{\tau_1, q_8\}$, $\{\tau_2, q_9\}$, $\{\tau_3, q_{10}\}$ and $\Delta\simeq2^{55}$ (see Table~\ref{tab:prime-system}).
Then, the set of primes used at level 8 or higher will include all the terminal primes.
In this way, we can guarantee the top-level ciphertext modulus is equal to $PQ_\text{max}$.
The actual implementation in \NAME allows diverse $\Delta$ values for \boot levels through a complex scale management with non-\bitprime{30} $q_i$'s and delayed rescaling.

\subsection{Inverted-Terminal Data Layout}
\label{sec:32bit:layout}

We prefer to allocate each polynomial in a contiguous and aligned memory address space to simplify index calculation in GPU kernels and to reduce the overhead of memory allocation.
The use of rational rescaling complicates this because the limbs corresponding to the same RNS prime can be placed at different memory positions for each level.

We propose an inverted-terminal data layout to solve this problem.
Taking level 3 ($\Delta\simeq2^{40}$) as an example, at which we utilize four \bitprime{30} ($q_0, q_1, q_2, q_3$) and two \bitprime{25} ($\tau_0, \tau_1$), we place the limbs of a polynomial in the order of [$\tau_1, \tau_0, q_0, q_1, q_2, q_3$] (see Fig.~\ref{fig:rescaling}).
Because we select $\tau_i$'s and $q_i$'s in a fixed order (\S\ref{sec:32bit:prime-system}), we can guarantee that polynomial data are allocated contiguously in memory.
The simplified index calculation from our inverter-terminal data layout is especially beneficial for computations between polynomials at different levels, such as in rescaling (see Fig.~\ref{fig:sequential-fusion}).

\section{Operation-Level Optimizations}
\label{sec:kernel}

\subsection{32-bit Modular Arithmetic}
\label{sec:kernel:montgomery}

We implement signed Montgomery reduction (SMR)~\cite{iacr-2018-signed-montgomery} for GPUs.
Alg.~\ref{alg:signed} shows its computation process.
While prior studies utilized classical reduction methods from Barrett~\cite{eurocrypt-1986-barrett}, Montgomery~\cite{1985-montgomery}, and Shoup~\cite{shoup-ntl}, Table~\ref{tab:reduction} shows that SMR is cheaper than the classical methods.
SMR includes fewer instructions than Barrett or Montgomery reduction.
Shoup reduction requires slightly less computation than SMR but Shoup reduction is restricted to constant mult and requires extra memory access for precomputed constants.
Therefore, we use SMR for all modular mult in \NAME.

Potential drawbacks of using SMR are additional computation requirements to convert to or from the Montgomery form; in the Montgomery form, each data element $x$ of a polynomial is stored as $x \cdot 2^{32} \text{ mod } q_i$.
We completely eliminate this overhead by kernel fusion in \S\ref{sec:seq:sequential}.

\subsection{Kernel Optimizations}
\label{sec:kernel:specific}

\begin{algorithm}[tb!]
\small
\caption{32-bit signed Montgomery reduction (SMR)}
\label{alg:signed}
\begin{algorithmic}[1]
\Require $x \in [-q_i \cdot 2^{31}, q_i \cdot 2^{31})$, $q_i < 2^{31}$, $m_i \in [-2^{31}, 2^{31})$, $m_i = q_i^{-1} \text{ mod } 2^{32}$ \Comment{{\footnotesize Constant $m_i$ is precomputed for each $q_i$.}}
\Ensure $y = x \cdot 2^{-32} \text{ mod } q_i$, $y \in (-q_i, q_i)$
\State $x = x_\text{hi} \cdot 2^{32} + x_\text{lo}$, $x_\text{lo} \in [0, 2^{32})$ \Comment{{\footnotesize Bit extraction}}
\State $z \gets \text{mullo32}(x_\text{lo}, m_i)$ \Comment{{\footnotesize 32-bit signed mult (retains lower 32 bits)}}
\State $z \gets \text{mulhi32}(z, q_i)$ \Comment{{\footnotesize 32-bit signed mult (retains upper 32 bits)}}
\State $y \gets x_\text{hi} - z$
\end{algorithmic}
\end{algorithm}

\setlength{\tabcolsep}{2pt}
\begin{table}
\small
\centering
\caption{Comparison of modular reduction methods. On recent GPU architectures, the computational cost is in the order of $\text{mulwide} \ge \text{mulhi} > \text{mullo}$. We ignore signedness. \# of const refers to the number of precomputed constants required per RNS prime for modular reduction.}
\label{tab:reduction}
\vspace{-0.04in}
\begin{tabularx}{0.99\columnwidth}{l|Lcc}
\toprule
Reduction & \multicolumn{1}{c}{Computation} & \# of & Output \\
 method       & \multicolumn{1}{c}{requirements} & const & range\\
\midrule
Barrett & mulhi64 + mullo64 + add64 & 1 & $[0, 2q_i)$\\
Montgomery & mulwide32\textsuperscript{\dag} + mullo32 + add64 & 1 & $[0, 2q_i)$\\
Shoup & 2 * mullo32 + add32 & Many\textsuperscript{\ddag} & $[0, 2q_i)$\\
Signed Mont. & mulhi32 + mullo32 + add32 & 1 & $(-q_i, q_i)$ \\
\bottomrule
\end{tabularx}
\begin{itemize}
    \footnotesize
    \item[\textsuperscript{\dag}] mulwide32 multiplies two 32-bit integers and produces a 64-bit result.
    \item[\textsuperscript{\ddag}] Shoup reduction is used only for constant mult, and requires an additional precomputed constant for every unique constant to multiply.
\end{itemize}
\vspace{-0.04in}
\end{table}
\setlength{\tabcolsep}{6pt}

\begin{figure}
    \centering
    \subfloat{\label{fig:roofline}}
    \subfloat{\label{fig:hrot-breakdown}}
    \includegraphics[width=0.98\linewidth]{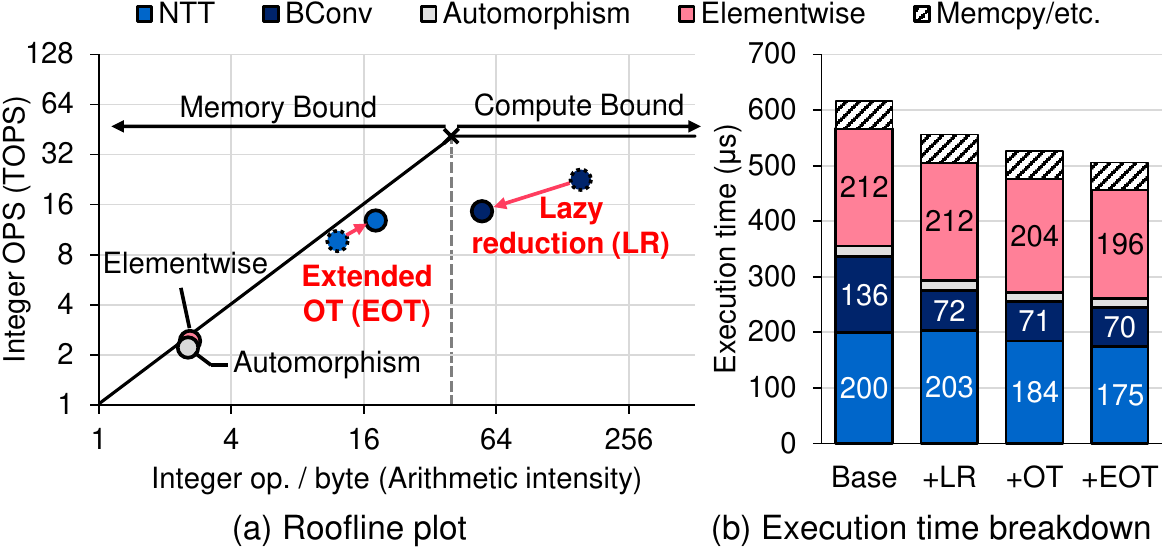}
    \Description{}
    \caption{(a) Roofline plot and (b) execution time breakdown of \hrot with incremental optimizations---lazy reduction (LR), on-the-fly twiddle factor generation (OT)~\cite{iiswc-2020-ntt}, and extended OT (EOT). \hrot is performed for a ciphertext with 48 limbs on an RTX 4090 GPU. $N=2^{16}$. $\alpha=12$.}
    \label{fig:roofline-and-breakdown}
\end{figure}

The roofline plot in Fig.~\ref{fig:roofline} summarizes the computational characteristics of polynomial operations comprising \hrot.
Automorphism and element-wise operations involve little amount of computation and are bottlenecked by limited GPU memory (DRAM or L2 cache) bandwidth.
Their inherently simple structure hinders further kernel-level optimization.
Meanwhile, on RTX 4090, BConv is mostly compute-bound and (I)NTT is mostly memory-bound.
We devise optimization methods addressing these characteristics.

\subsubsection*{BConv matrix mult with lazy reduction}

We implement BConv matrix mult in Alg.~\ref{alg:modsw} with a well-known blocked matrix-matrix mult method~\cite{cutlass-blog}.
However, unlike regular matrix-matrix mult, BConv matrix mult includes modular reductions, which introduce significant compute overhead.

To reduce the number of modular reductions, we delay the reduction, which is a widely used technique (e.g., 100$\times$~\cite{tches-2021-100x}) referred to as lazy reduction.
However, lazy reduction often requires restricting the range of RNS primes (e.g., $q_i\in[0, 2^{30})$~\cite{jsc-2014-harvey-ntt}) to exploit extra bits in a word.

We discover that, even with the minimal prime restriction ($q_i\in[0, 2^{31})$) in \NAME, SMR allows sufficiently lazy reduction for BConv.
We prepare $temp[j]$ and $(Q/q_j)^{-1} \text{ mod } p_i$ (see Alg.~\ref{alg:modsw}) respectively in the ranges of $(-q_j/2, q_j/2) \in (-2^{30}, 2^{30})$ and $(-p_i/2, p_i/2)$, whose product will be in the range of $(-2^{29}\cdot p_i, 2^{29}\cdot p_i) \in (-2^{60}, 2^{60})$.
Thus, we can accumulate up to eight products, which will be in the range of $(-2^{32}\cdot p_i, 2^{32}\cdot p_i)$, without an overflow for a signed 64-bit integer.
For more accumulation, we add or subtract $2^{32}\cdot p_i$ to adjust the range to $(-2^{31}\cdot p_i, 2^{31}\cdot p_i)$, after which we can accumulate four more products.
We repeat this until we accumulate all the products.
Then, we use SMR for the final reduction.
For regular unsigned Montgomery reduction, the product will be in a twice wider range $[0, 2^{31}\cdot p_i)$, requiring the range adjustment to be performed twice as much.

\begin{figure*}
    \centering
    \includegraphics[width=0.97\linewidth]{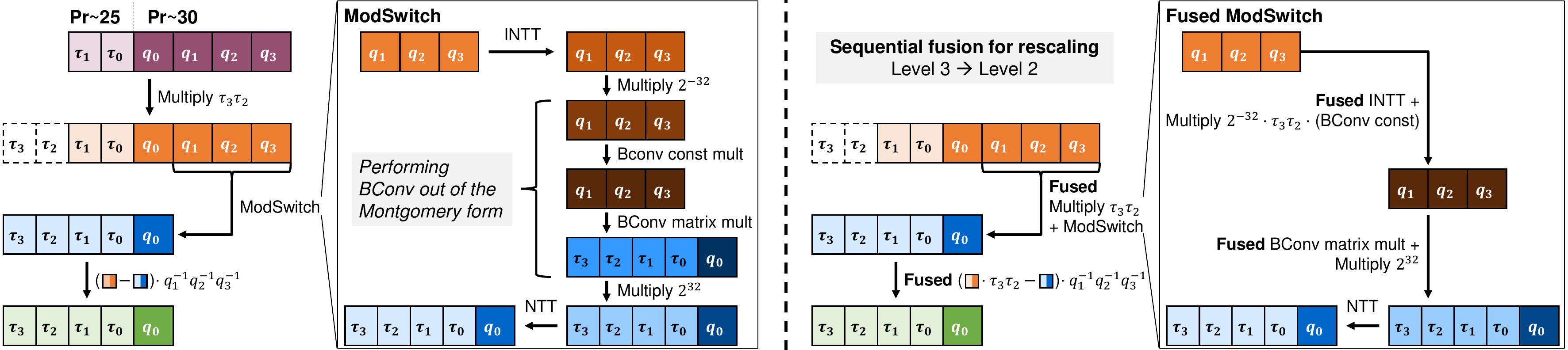}
    \vspace{-0.02in}
    \Description{}
    \caption{Operational sequence of rescaling from level 3 to level 2 ($\Delta=2^{40}$) without (left) and with sequential fusion (right). Each block represents a limb of a polynomial. Each arrow represents an operation implemented with one or two GPU kernels.}
    \label{fig:sequential-fusion}
    \vspace{-0.02in}
\end{figure*}

\begin{figure}
    \centering
    \includegraphics[width=0.95\linewidth]{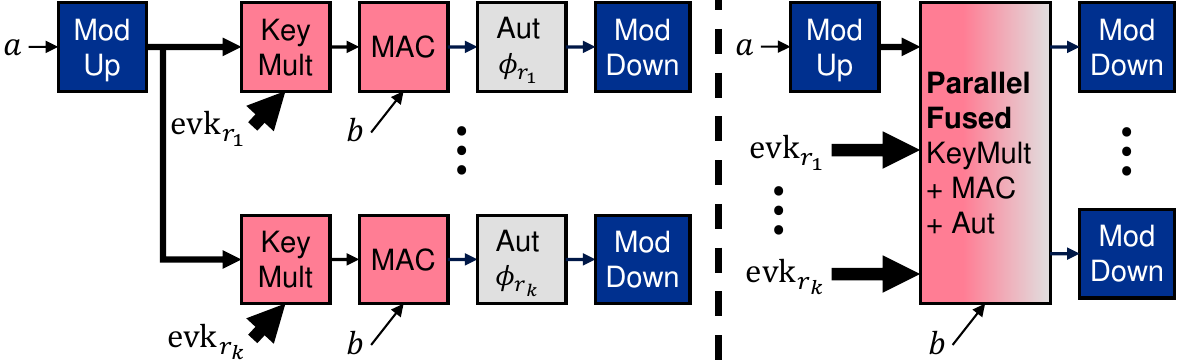}
    \vspace{-0.03in}
    \Description{}
    \caption{Part of FHE linear transform operational sequence without (left) and with (right) parallel fusion. Blocks represent operations (red: element-wise, gray: automorphism, blue: ModSwitch variant). Arrows show dataflow and arrow thickness represents the size of data.}
    \label{fig:parallel-fusion}
    \vspace{-0.02in}
\end{figure}

\subsubsection*{(I)NTT with extended on-the-fly twiddle factor generation (EOT)}
We use the standard radix-2 Cooley-Tukey~\cite{cooley-tukey} FFT, which involves $\log_2 N$ stages of butterfly operations, to implement (I)NTT kernels.
We partition the stages into a small number of phases, where synchronization between threads are required in between the phases.
At each stage, data are multiplied by constants named twiddle factors.
As each limb requires a unique set of $N$ twiddle factors, whose total size is the same as the limb itself, twiddle factors become significant sources of memory access.

We can use on-the-fly twiddle factor generation (OT)~\cite{iiswc-2020-ntt} to reduce the memory burden.
OT generates a twiddle factor $\overline{W}=W^{k}$ during runtime by computing $\overline{W}=\overline{W}_\text{lo}\cdot\overline{W}_\text{hi}$, where $\overline{W}_\text{lo}=W^{\text{lsb}(k)}$, $\overline{W}_\text{hi}=W^{k - \text{lsb}(k)}$, and lsb$(\cdot)$ is a function extracting a number of lower bits from an integer.

While Kim et al.~\cite{iiswc-2020-ntt} only used OT for the last stage of each phase, we propose extended OT (EOT), which iteratively generates all the twiddle factors for every stage in a phase.
We give an example for processing three stages in a phase.
Each thread needs twiddle factors $\{W^{\text{brev}(i)}\}$ for the first stage, $\{W^{\text{brev}(2 \cdot i)}, W^{\text{brev}(2 \cdot i + 1)}\}$ for the second, and $\{W^{\text{brev}(4 \cdot i)}, W^{\text{brev}(4 \cdot i + 1)}, W^{\text{brev}(4 \cdot i + 2)}, W^{\text{brev}(4 \cdot i + 3)}\}$ for the last, where brev refers to bit-reversal.
We first generate the last four twiddle factors by OT.
Then, EOT iteratively generates the twiddle factors for the preceding stages as follows:
\begin{align*}
&W^{\text{brev}(2 \cdot i)} \gets (W^{\text{brev}(4 \cdot i)})^2 \text{ \& } W^{\text{brev}(2 \cdot i + 1)} \gets (W^{\text{brev}(4 \cdot i + 2)})^2,\\
&W^{\text{brev}(i)} \gets (W^{\text{brev}(2 \cdot i)})^2.
\end{align*}

\subsubsection*{Optimization effects}

Fig.~\ref{fig:hrot-breakdown} shows the performance improvements achieved for \hrot by applying the optimizations above.
Our SMR-based lazy reduction (+LR) eliminates 65\% of the integer operations in BConv matrix mult kernels, which leads to a 1.89$\times$ speedup of BConv in \hrot.
Also, EOT reduces the amount of twiddle factor load by 96\% and requires less cache space, which also indirectly enhances the performance of other memory-bound kernels.
Overall, EOT achieves a 1.16$\times$ speedup of NTT for \hrot, compared to the 1.10$\times$ speedup of regular OT~\cite{iiswc-2020-ntt}.

Fig.~\ref{fig:roofline} shows that the lazy reduction and EOT respectively make BConv and (I)NTT kernels more balanced in terms of arithmetic intensity on RTX 4090.
However, different GPU architectures create different trade-offs, possibly demanding alternative approaches to implement the kernels.

\subsubsection*{Fine-tuning}

Thus, we implement the kernels in a highly parameterized manner such that we can easily modify how much job each thread will take charge, how the (I)NTT stages will be partitioned into multiple phases, to what extent EOT will be applied, what block sizes will be used for the BConv matrix mult kernel based on the blocked matrix-matrix mult method~\cite{cutlass-blog}, and so on.
We fine-tune the kernels for each GPU architecture using our software fine-tuner (\S\ref{sec:impl}).

\section{Operational Sequence Optimization}
\label{sec:seq}

Our optimized 32-bit implementation effectively accelerates compute-intensive BConv and (I)NTT operations.
However, element-wise operations and automorphism, which have low arithmetic intensity and are bound by memory bandwidth, do not get improved simply by operation-level optimizations.
They become serious sources of performance bottlenecks (see Fig.~\ref{fig:hrot-breakdown}) with the aforementioned optimizations reducing the significance of (I)NTT and BConv.

We thoroughly analyze common operational sequences in FHE and find opportunities to reduce the memory overhead induced by these operations.
Our main tool is \emph{kernel fusion}.
we discovered new fusion opportunities by analyzing the modified operation sequences that arise from rational rescaling with our 25-30 prime system and signed Montgomery reduction.
Additionally, we apply fusion across a wider range of degrees rather than just a few basic sequences.
Fusion techniques can be categorized into two types: sequential fusion and parallel fusion. To clearly distinguish between these two categories, we use \seqf{X} and \parf{X} notations.

Prior GPU implementations of FHE, including 100$\times$~\cite{tches-2021-100x} and WarpDrive~\cite{hpca-2025-warpdrive}, have also utilized kernel fusion but in a limited form.
100$\times$ fuses BConv const mult in ModSwitch (lines 3--4 in Alg.~\ref{alg:modsw}) with the preceding INTT (\seqf{1}). 
Additionally, 100$\times$ implements KeyMult (\parf{1}), and batched \pmult (\parf{2}) in single kernels, respectively.
WarpDrive further applied fusion to batching independent polynomial operations (\parf{3}).
\NAME incorporates all of these existing fusion techniques and extends them.

\vspace{-0.05in}
\subsection{Sequential Fusion}
\label{sec:seq:sequential}

We first find opportunities at the level of basic FHE mechanisms, such as \hmult, \hrot, and rescaling.
We discover methods to fuse kernels for element-wise operations or automorphism forming a sequential dataflow in these mechanisms.
We use an example of rescaling from level 3 to level 2 when $\Delta=2^{40}$ (see Fig.~\ref{fig:sequential-fusion}).

For rescaling, we first multiply $\tau_3\tau_2$, which are primes that are used at level 2 but not at level 3, to the polynomial to temporarily increase the modulus.
Then, we want to divide it by $q_1q_2q_3$, but we first need to zero-out the residues corresponding to these primes.
Thus, we perform ModSwitch from $q_1q_2q_3$ to $\tau_3\tau_2\tau_1\tau_1q_0$ and subtract the result to the polynomial, which allows us to finally multiply $q_1^{-1}q_2^{-1}q_3^{-1}$ for the division.
Fig.~\ref{fig:sequential-fusion} shows that this process requires a lot of constant mults.
In addition to the aforementioned constants, we also multiply constants for the transitions to/from the Montgomery form ($2^{32}, 2^{-32}$) and the BConv const.

We accelerate this sequence by fusing constant mults into adjacent INTT and BConv kernels (\seqf{2}).
Our approach is distinct from conventional kernel fusion approaches~\cite{tches-2021-100x} in that the fused INTT or BConv kernels do not have any additional operations due to the precomputation of the constants.
INTT kernels already include constant mult by $N^{-1}\cdot2^{32}$ (Montgomery form) at the end; we can instead prepare $N^{-1}\cdot\tau_3\tau_2\cdot(\text{BConv const})$ to perform the fusion.
Also, we can multiply $2^{32}$ to the BConv matrix in advance.
The use of precomputed constants reduces overall computation, notably eliminating the cost of Montgomery conversion entirely.

Also, we split (fiss) and reorder operations~\cite{mlsys-2019-fusion} to find more fusion opportunities (\seqf{3}).
For example, we split $\tau_3\tau_2$ mult into two: one for $\{\tau_1,\tau_0,q_0\}$-limbs and the other for $\{q_1,q_2,q_3\}$-limbs.
We reorder the operational sequence to merge the former with the last kernel of rescaling and the latter with INTT, as illustrated in Fig.~\ref{fig:sequential-fusion}. 

We apply such techniques for sequential fusion to various CKKS mechanisms.
For example, \hmult and \hrot also include a similar operational sequence (ModDown~\cite{rsa-2020-better}), where we can perform the same optimization.

\subsection{Parallel Fusion}
\label{sec:seq:parallel}

The most basic form of parallel fusion is batching. 
We apply batching to multiple polynomial-wise operations, similar to \parf{3} of WarpDrive.
For example, we merge two or three polynomial addition kernels into a single kernel for the \hadd mechanism.
TensorFHE~\cite{hpca-2023-tensorfhe} goes even further by batching multiple mechanism evaluations, which is possible when the server can batch requests from multiple queries.
However, as we find this server operation model limited to few use cases, considering the currently high latency of FHE workloads, we focus on the basic single-query model.
Also, batching in itself does not reduce memory access without shared data.

Next, we apply parallel fusion for sequences that involve accumulations (\parf{4}).
Applying parallel fusion for such sequences can significantly reduce memory access as we remove the need to store or load temporary accumulation results.
Many FHE mechanisms, ranging from basic \hmult/\hrot to high-level polynomial evaluations, have such sequences.
Especially, by reordering the operational sequences, we discover sequences that accumulate multiple automorphism results, such as $\phi_{r_1}(a_1) + \cdots + \phi_{r_k}(a_k)$.
We implement parallel fusion for this sequence by using the inverse of Eq.~\ref{eq:automorphism} ($\phi_r^{-1}$) to compute this in a single kernel.
Each thread can efficiently accumulate multiple automorphism results into a single register as
\begin{equation}
    a_\text{res}[j][i] = a_1[j][\phi_{r_1}^{-1}(i)] + \cdots + a_k[j][\phi_{r_k}^{-1}(i)].
\end{equation}
We also fuse sequences that compute multiple accumulation results with the same input, which are prevalent in \boot.

Finally, we apply parallel fusion for high-level linear transform~\cite{eurocrypt-2021-efficient} and other DNN mechanisms~\cite{usenixsec-2018-gazelle, icml-2022-resnet, privatenlp-2020-rnn} (\parf{5}).
Linear transforms require a number of parallel \hrot operations, and \cite{eurocrypt-2021-efficient} proposed methods to reuse common computational results within these sequences.
As an example, we show an optimized linear transform sequence in Fig.~\ref{fig:parallel-fusion}.
In this case, rather than simply batching operations in parallel, we focus on reducing redundant memory access for shared inputs.
We merge multiple parallel element-wise (KeyMult and MAC) and automorphism operations into a single kernel, significantly reducing memory access to shared inputs such as $\text{ModUp}(a)$ and $b$.

\section{Implementation}
\label{sec:impl}

\begin{figure}
    \centering
    \includegraphics[width=0.97\columnwidth]{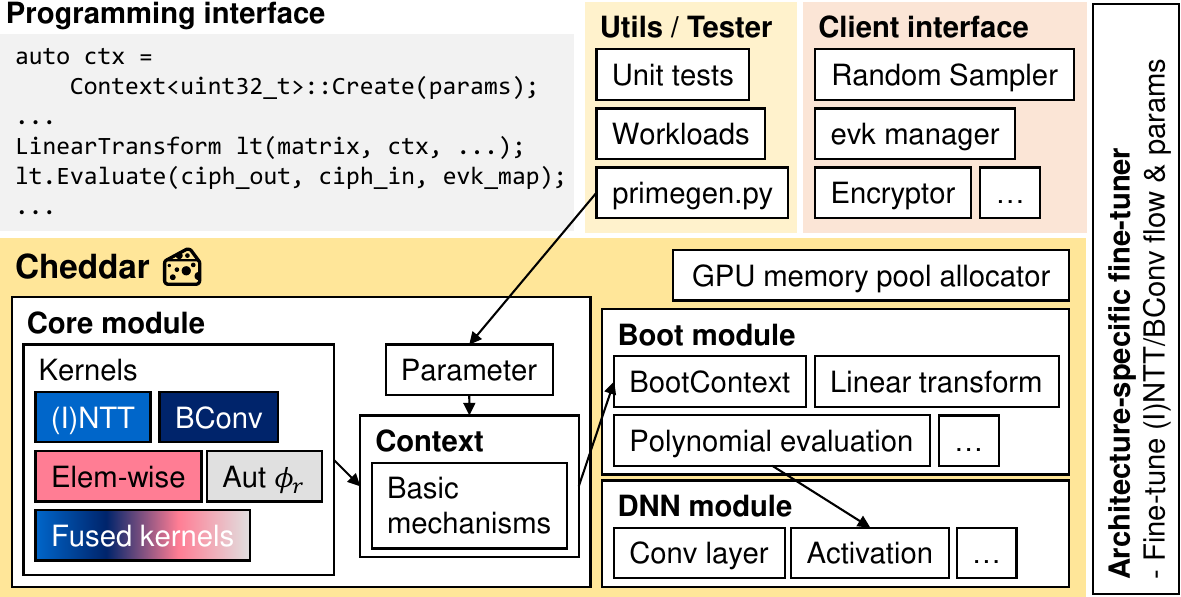}
    \vspace{-0.07in}
    \Description{}
    \caption{Software implementation of \NAME.}
    \label{fig:impl}
    \vspace{-0.09in}
\end{figure}

Fig.~\ref{fig:impl} shows the overall software structure of \NAME.
The core module of \NAME contains the GPU kernels and implementations of the basic mechanisms.
\boot and DNN modules, which are optional, provide implementations of various common high-level mechanisms, such as linear transform, arbitrary polynomial evaluation, and convolutional layers.
The aforementioned kernel and operational sequence optimizations are fully applied to each mechanism.
The main \NAME library is implemented in 11,000+ lines of C++/CUDA code.
Clients can easily utilize the library with our high-level C++ programming interface with minimal low-level knowledge.

We provide an architecture-specific fine-tuner to determine the dataflow and parameters for the highly parameterized GPU kernels discussed in \S\ref{sec:kernel:specific}.
The fine-tuner performs a comprehensive design space exploration to find the best configuration for each GPU architecture.
We also provide various utilities, such as a script (primegen.py) for our 32-bit RNS construction (\S\ref{sec:32bit:prime-system}), and a test-purpose client interface.

\NAME incorporates state-of-the-art FHE algorithms.
We used fast high-precision \boot algorithms~\cite{acns-2022-sparseboot, eurocrypt-2021-efficient, rsa-2020-better} and extended the ModDown merging optimization from MAD~\cite{micro-2023-mad} to support our prime system.
Also, we removed a portion of \hrot evaluations during \boot by the pre-rotation removal technique in ARK~\cite{micro-2022-ark}.
We further devised a method to modify the linear transform matrices in \boot to reduce the amount of \hconj (mechanism similar to \hrot used for conjugation) evaluations.
We also applied workload-specific optimizations that have profound performance implications.
For example, we merged \boot for $[\!\langle\Delta\cdot\mathbf{u}\rangle\!]$ and $[\!\langle\Delta\cdot\mathbf{v}\rangle\!]$ encrypting real-valued $\mathbf{u}$ and $\mathbf{v}$ into a single \boot of $[\!\langle\Delta\cdot(\mathbf{u} + \sqrt{-1}\cdot\mathbf{v})\rangle\!]$, which halves the \boot cost
for our HELR implementation in \S\ref{sec:eval}.

\section{Evaluation}
\label{sec:eval}

\subsection{Experimental Setup}
\label{sec:eval:setup}

By default, we used 128-bit secure parameters with $N=2^{16}$, $PQ<2^{1776}$, and $\Delta=2^{40}$.
We set $\mathtt{dnum}$~\cite{rsa-2020-better} as four.
For evaluation, we implemented the following representative FHE CKKS workloads widely used in prior work.

\begin{itemize}
    \item \textbf{bts}: Full-slot bootstrapping of a ciphertext encrypting a length-$2^{15}$ complex vector. $L_\text{eff}=13$ for $\Delta=2^{40}$. We used higher scales of $\Delta=2^{55}\text{--}2^{58}$ during \boot.
    \item \textbf{HELR}~\cite{aaai-2019-helr}: Logistic regression binary classification model training. We performed 32 training iterations, each with a 1024-batch of $14\!\times\!14$ grayscale images. We use execution time per iteration (ms/it) for comparison. 
    \item \textbf{ResNet}~\cite{icml-2022-resnet}: ResNet-20 CNN inference on a single CIFAR-10 image of size $32\times 32 \times 3$. Each ReLU uses 14 levels.
    \item \textbf{Sort}~\cite{tifs-2021-sorting}: Two-way sorting network with $2^{14}$ numbers.
\end{itemize}

\setlength{\tabcolsep}{4pt}
\begin{table}[t]
\centering
\small
\caption{Peak 32-bit integer math throughput and DRAM configuration of GPUs based on official specifications. We used the six GPUs listed below for the evaluation of \NAME.}
\vspace{-0.06in}
\label{tab:gpu}
\begin{tabular}{l|rr}
  \toprule
  GPU & Int32 throughput & DRAM (capacity, bandwidth) \\
  \midrule
  V100 & 14.1 TOPS & HBM2 (16GB, {\color{white}0}897GB/s)\\
  A100 40GB & 19.5 TOPS & HBM2 (40GB, 1555GB/s) \\
  A100 80GB & 19.5 TOPS & HBM2E (80GB, 1935GB/s)\\
  H100 & 25.6 TOPS & HBM2E (80GB, 2039GB/s)\\
  RTX 4090 & 41.3 TOPS & GDDR6X (24GB, 1008GB/s)\\
  RTX 5090 & 104.8 TOPS & GDDR7 (32GB, 1792GB/s)\\
  \midrule
  MI100 & 23.1 TOPS & HBM2 (32GB, 1229GB/s) \\
  \bottomrule
\end{tabular}
\end{table}
\setlength{\tabcolsep}{6pt}

We used various GPUs in Table~\ref{tab:gpu} to compare \NAME with prior state-of-the-art GPU implementations on the same GPU device: V100, A100 40GB, or A100 80GB.
A100 80GB has 1.24$\times$ higher DRAM bandwidth compared to A100 40GB as well as having larger DRAM capacity.
For the comparison with AMD MI100 GPU, we used A100 40GB for \NAME as it shows similar compute and memory characteristics.
We also tested \NAME on more recent GPU devices, including a server GPU (H100) and a consumer GPU (RTX 5090).

\NAME also supports 64-bit execution through a poly-morphic code design.
Further, we can use various RNS constructions using single or double rescaling, other than rational rescaling with our 25-30 prime system.
We use those suboptimal configurations of \NAME for ablation studies.

FHE workloads require large memory capacity to hold various \evks, ciphertexts, and other polynomials.
Due to the out-of-memory (OoM) errors, we could not run some workloads on V100 or RTX 4090 with relatively small memories.

For aggressive parameter sets with $\Delta \le 2^{35}$, we use a sparse secret ($h=256$)~\cite{korean-2022-cheon-practical} for ResNet and HELR, which still ensures 128-bit security.

\subsection{FHE Workload Performance}
\label{sec:eval:workload}

\setlength{\tabcolsep}{2pt}
\begin{table}[t]
\centering
\small
\caption{Execution time of FHE CKKS workloads using \NAME compared to 100$\times$~\cite{tches-2021-100x}, TensorFHE~\cite{hpca-2023-tensorfhe}, HEaaN-GPU~\cite{heaan-latest, discc-2024-fhe-cnn}, WarpDrive~\cite{hpca-2025-warpdrive}, GME~\cite{micro-2023-gme}, FAB~\cite{hpca-2023-fab}, Poseidon~\cite{hpca-2023-poseidon}, and EFFACT~\cite{hpca-2025-effact}. OoM refers to execution failure due to the out-of-memory error.}
\vspace{-0.05in}
\label{tab:workload-performance}
    \begin{tabularx}{0.99\columnwidth}{l|RRR}
    \toprule
    \multicolumn{1}{c|}{Implementation} & \multicolumn{1}{c}{\boot} & \multicolumn{1}{c}{HELR\textsuperscript{\dag}} & \multicolumn{1}{c}{ResNet}\\
     \multicolumn{1}{c|}{(Hardware)} & \multicolumn{1}{c}{(ms)} & \multicolumn{1}{c}{(ms/it)} & \multicolumn{1}{c}{(s)} \\
    \midrule
    100$\times$ (V100) & 328{\color{white}.0} & 775{\color{white}.0}  & \centerdashtwo\\
    \quad vs. \NAME (V100) & 4.44$\times$ & 9.74$\times$\\
    TensorFHE (A100 40GB)* & 250{\color{white}.0} & 1007{\color{white}.0} & 4.94\\
    \quad vs. \NAME (A100 40GB) & 5.88$\times$ & 19.6$\times$ & 3.63$\times$\\
    GME-base (MI100) & 413{\color{white}.0} & 658{\color{white}.0}  & 9.99\\
    \quad vs. \NAME (A100 40GB) & 9.72$\times$ & 12.8$\times$ & 7.35$\times$\\
    HEaaN-GPU (A100 80GB) & 171{\color{white}.0} & \centerdashtwo  & 8.58\\ 
    \quad vs. \NAME (A100 80GB) & 4.28$\times$ & & 6.50$\times$\\
    WarpDrive (A100 80GB) &121{\color{white}.0} & 113{\color{white}.0} & 5.88\\
    \quad vs. \NAME (A100 80GB) & 3.03$\times$ & 2.18$\times$ & 4.45$\times$\\
    \midrule
    \NAME (V100) & 73.8 & 79.6  & OoM \\
    \NAME (A100 40GB) & 42.5 & 51.4  & 1.36 \\
    \NAME (A100 80GB) & 40.0 & 51.9  & 1.32 \\
    \NAME (H100) & 31.2 & 40.7  & 1.05 \\
    \NAME (RTX 4090) & 31.6 & 29.9  & OoM \\
    \NAME (RTX 5090) & 22.1 & 25.9 & 0.72\\
    \midrule
    \NAME (H100, $\Delta=2^{35}$) & 28.7 & 39.3 & 0.82 \\
    \NAME (RTX 5090, $\Delta=2^{35}$) & 19.8 & 25.5 & 0.57 \\
    \midrule
    FAB (FPGA) & 477{\color{white}.0} & 103{\color{white}.0}  & \centerdash \\
    Poseidon (FPGA) & 128{\color{white}.0} & 72.9  & 2.66\\
    EFFACT (FPGA) & 148{\color{white}.4} & 64.6  & 2.18\\
    GME (Modified MI100) & 33.6 & 54.5 & 0.98\\
    \bottomrule
    \end{tabularx}
    \begin{itemize}
        \footnotesize
        \item[*] TensorFHE batched workload evaluations on multiple input ciphertexts. We divided the execution time with the batch size in favor of TensorFHE.
        \item[\textsuperscript{\dag}] HELR includes merged \boot.
        Excluding this optimization shows 57.4 ms/it on A100 80GB, reducing the speedup of \NAME over WarpDrive from 2.18$\times$ to 1.96$\times$.
    \end{itemize}
\vspace{-0.09in}
\end{table}
\setlength{\tabcolsep}{6pt}

\NAME achieves speedups ranging from 2.18$\times$ to 19.6$\times$ for the workloads compared to the state-of-the-art GPU implementations~\cite{tches-2021-100x, hpca-2023-tensorfhe, heaan-latest, hpca-2025-warpdrive, discc-2024-fhe-cnn, micro-2023-gme}.
In particular, \NAME shows 2.18--4.45$\times$ improved performance compared to WarpDrive~\cite{hpca-2025-warpdrive}, which is one of the latest GPU implementations, when comparing the two on the same A100 80GB GPU.

\NAME also outperforms HEaaN-GPU~\cite{heaan-latest,discc-2024-fhe-cnn}, achieving 6.50$\times$ faster ResNet inference time of 1.32 seconds.
Meanwhile, HEaaN-GPU also reported an inference time of 1.40 seconds by using a modified FHE CNN implementation based on HyPHEN~\cite{access-2024-hyphen} and AESPA~\cite{arxiv-2022-aespa}, which replaces the 14-level ReLU with a 1-level activation.
Such direct workload enhancements can also benefit \NAME but we did not adopt them for fair comparison with other GPU implementations.

Using \NAME with more recent NVIDIA GPUs, such as the H100, RTX 4090, and the latest RTX 5090, resulted in even faster execution of the workloads.
Compared to the A100 80GB where \boot takes 40ms, the H100 and RTX 4090 reduce the time to around 31ms, and this is further accelerated to 22.1ms on an RTX 5090; it is faster to perform \boot than to receive a new 5MB ciphertext through a 1Gbps network ($>\!40$ms).
Further, with an H100 GPU, the encrypted ResNet inference latency is reduced to a practical level of 1.05s, or 0.82s if we opt for a more aggressive parameter choice with $\Delta=2^{35}$.
On an RTX 5090, this latency improves to 0.72s and 0.57s, respectively, because its int32 throughput of 104.8 TOPS is over 4$\times$ higher than the H100's 25.6 TOPS, despite its lower DRAM bandwidth (see Table~\ref{tab:scale_change}).

\NAME also demonstrates superior performance compared to custom hardware designs based on FPGA~\cite{hpca-2023-fab, hpca-2023-poseidon, hpca-2025-effact}.
\NAME proves to be a cost-effective solution for FHE acceleration given that using \NAME with a single RTX 4090 GPU (\$1,600) or RTX 5090 GPU (\$2,000) is substantially faster than the FPGA implementations, which used expensive Alveo U280 FPGAs (\$6,000).
\NAME also achieves similar performance to GME~\cite{micro-2023-gme}, which added 186mm\textsuperscript{2} worth of specialized computational components into an AMD MI100 GPU to turn it into an FHE accelerator.

\subsection{Efficiency of 25-30 Prime System}
\label{sec:eval:rns}

\begin{figure}[t]
     \centering
     \subfloat[$(\text{\boot time})/L_\text{eff}$ \& \efflevel]{\centering\includegraphics[width=0.46\columnwidth]{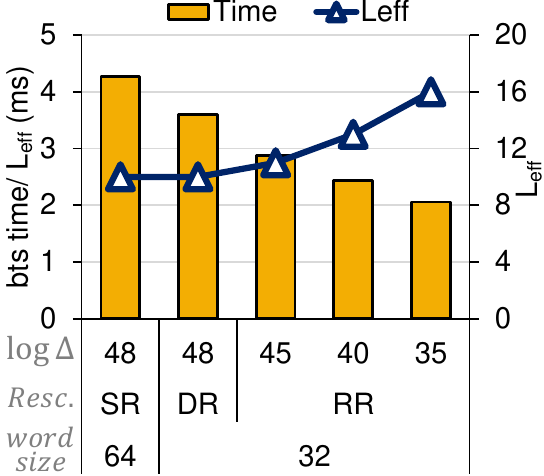}\label{fig:32bit-boot}}
     \hfill
     \subfloat[\boot portion \& \efflevel]{\includegraphics[width=0.52\columnwidth]{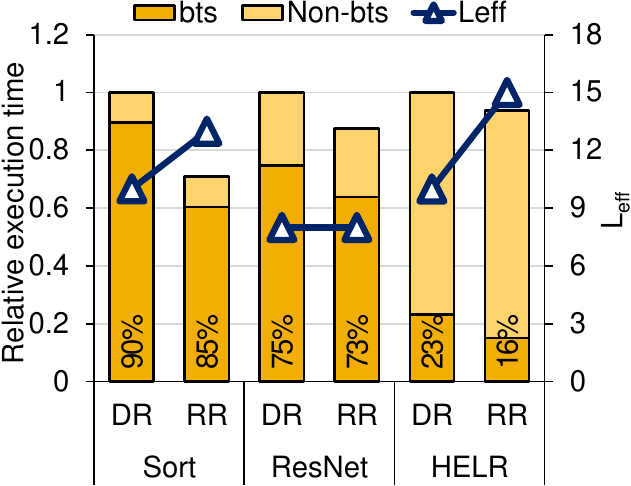}\label{fig:boot-ratio}}
     \vspace{-0.03in}
     \Description{}
     \caption{(a) \boot time divided by \efflevel and (b) relative execution time with \boot portion in the workloads depending on the RNS construction using single rescaling (SR), double rescaling (DR), or rational rescaling (RR) on H100.
     DR uses $\Delta=2^{48}$ and RR uses $\Delta=2^{40}$. Higher scales ($\Delta=2^{55}\text{--}2^{58}$) are used for \boot levels regardless of the configuration.}
     \label{fig:bench}
\end{figure}

\setlength{\tabcolsep}{2pt}
\begin{table}[t]
\centering
\small
\caption{Execution time (H100) and functionality of HELR,  ResNet, and Sort depending on $\Delta$. Higher scales ($\Delta=2^{55}\text{--}2^{58}$) are used for the \boot levels regardless of the configuration.}
\label{tab:scale_change}
    \vspace{-0.06in}
    \begin{tabularx}{0.99\columnwidth}{c|RR|RR|RR}
    \toprule
     & \multicolumn{2}{c|}{HELR} & \multicolumn{2}{c|}{ResNet} & \multicolumn{2}{c}{Sort} \\
    & \multicolumn{1}{c}{Time} & \multicolumn{1}{c|}{Acc.} & \multicolumn{1}{c}{Time} & \multicolumn{1}{c|}{Acc.} & \multicolumn{1}{c}{Time} & \multicolumn{1}{c}{Prec.\textsuperscript{\dag}}\\
    & \multicolumn{1}{c}{(ms/it)} & \multicolumn{1}{c|}{(\%)} & \multicolumn{1}{c}{(s)} & \multicolumn{1}{c|}{(\%)} & \multicolumn{1}{c}{(s)} & \multicolumn{1}{c}{(bits)} \\ 
    \midrule
    $\Delta=2^{48}$ (DR) & 44.3 & 96.42 & 1.20 & 91.96 & 14.83 & 12.84   \\ 
    $\Delta=2^{40}$ (RR) & 40.7 & 96.37 & 1.05 & 91.95 & 10.20 & 12.08 \\ 
    $\Delta=2^{35}$ (RR) & 39.3 & 96.12 & 0.82 & 91.88 & 9.51 & 11.90 \\
    $\Delta=2^{30}$ (SR) & 34.4 & 52.21 & 0.68 & 9.59 & 8.49 & 9.17 \\
    \midrule
    Unencrypted* & 3.5 & 96.37 & 0.003 & 92.52 & 0.45 & $\infty$\\
    \bottomrule
    \end{tabularx}
    \begin{itemize}
        \footnotesize
        \item[\dag] Average $\log_2\epsilon^{-1}$ (bits). $\epsilon = ||\mathbf{u}^* - \mathbf{u}||_\infty$ for the true sorted vector $\mathbf{u}^*$ and the decrypted FHE Sort evaluation result $\mathbf{u}$.
        \item[*] Unencrypted execution times are measured on a single-threaded AMD EPYC 7452.
        Sort is implemented using bitonic sorting network.
    \end{itemize}
    \vspace{-0.07in}
\end{table}

By using rational rescaling (RR) with our 25-30 prime system, we can utilize a parameter set that provides a fitting precision value tailored to each workload by selecting proper scales ($\Delta$).
We can generally get a greater \efflevel (\S\ref{sec:background:ckks}) by using smaller scales with rational rescaling, which is not possible for na\"{i}ve 32-bit execution based on double rescaling (DR).
A greater \efflevel leads to less frequent \boot invocations and performance improvements, represented by the reduction in \boot time divided by \efflevel as we use smaller scales (see Fig.~\ref{fig:32bit-boot}).
For the workloads, Fig.~\ref{fig:boot-ratio} shows that our default $\Delta=2^{40}$ (RR) setting increases \efflevel by up to five levels, reduces the portion of \boot in the workloads by 2--8\%, and overall results in 1.07--1.41$\times$ speedups, compared to our hand-tuned $\Delta=2^{48}$ (DR) setting.

However, precautions are needed as reducing the scale may compromise the functionality of the workloads.
We tested the functionality of HELR, ResNet, and Sort while reducing the scale down to $\Delta=2^{30}$ (see Table~\ref{tab:scale_change}).
At $\Delta=2^{30}$, HELR and ResNet stop functioning correctly.
At $\Delta=2^{35}$, although the accuracy values slightly drop, HELR and ResNet are functional.
Sort functions correctly with reduced precision for such small scales.
Compared to the na\"{i}ve $\Delta=2^{48}$ (DR) setting, 1.13--1.56$\times$ speedups can be achieved for these workloads with the aggressive $\Delta=2^{35}$ (RR) setting.
We used $\mathtt{dnum}=6$~\cite{rsa-2020-better} for Sort with $\Delta=2^{48}$ and ResNet with $\Delta\ge2^{40}$ to abide by the 128-bit security requirements.

Finally, Fig.~\ref{fig:32bit-boot} also shows that even for equivalent settings (48-SR vs. 48-DR), 32-bit execution is much more efficient than 64-bit execution, reducing the \boot time by 1.18$\times$.

\vspace{-0.07in}
\subsection{FHE Mechanism Performance}
\label{sec:eval:operation}

\setlength{\tabcolsep}{2pt}
\begin{table}[t]
\centering
\small
\caption{Median execution times (RTX 4090) of \hmult, \hrot, \hadd, and rescaling when using \NAME or open-source GPU libraries~\cite{github-liberate_FHE, iacr-2024-heongpu, tdsc-2024-phantom}.
For the latter using 64-bit RNS, we halved the number of limbs and $\alpha$ for fair comparison.}
\vspace{-0.06in}
\label{tab:operation-performance}
     \begin{tabularx}{0.99\columnwidth}{l|c|c|RRRR}
    \toprule
    \multirow{2}{*}{Library} & \multirow{2}{*}{limbs} & \multirow{2}{*}{$\alpha$} & \multicolumn{4}{c}{Execution time ($\mu$s)} \\
     &  &  & \hmult & \hrot & \hadd & Rescale \\
    \midrule
    Liberate.FHE~\cite{github-liberate_FHE} & 24 & 6 & 5989 & 4554 & 96 & 150 \\
    HEonGPU~\cite{iacr-2024-heongpu} &24 & 6 & 2427 & 2279 & 79 & 255 \\
    Phantom~\cite{tdsc-2024-phantom} &24 & 6 & 958 & 850 & 83 & 114 \\
    \textbf{\NAME} &48 & 12 & 533 & 476 & 48 & 68\\
    \midrule
    Liberate.FHE~\cite{github-liberate_FHE} &12 & 6 & 2848 & 2052 & 22 & 120 \\
    HEonGPU~\cite{iacr-2024-heongpu} & 12 & 6 & 825 & 740 & 10 & 113 \\
    Phantom~\cite{tdsc-2024-phantom} & 12 & 6 & 380 & 336 & 18 & 68 \\
    \textbf{\NAME} & 24 & 12 & 222 & 191 & 11 & 47\\
    \bottomrule
    \end{tabularx}
    \vspace{-0.12in}
\end{table}

We used three open-source GPU FHE libraries, Phantom~\cite{tdsc-2024-phantom}, HEonGPU~\cite{iacr-2024-heongpu}, and Liberate.FHE~\cite{github-liberate_FHE}, to compare the performance of basic mechanisms with \NAME.
By adjusting the number of limbs and $\alpha$, we made sure all the libraries operate with the same amount of input data, excluding the performance gains stemming from our 25-30 prime system.
Table~\ref{tab:operation-performance} shows the results on an RTX 4090 GPU.

\begin{figure*}[ht]
     \centering
     \includegraphics[width=0.99\textwidth]{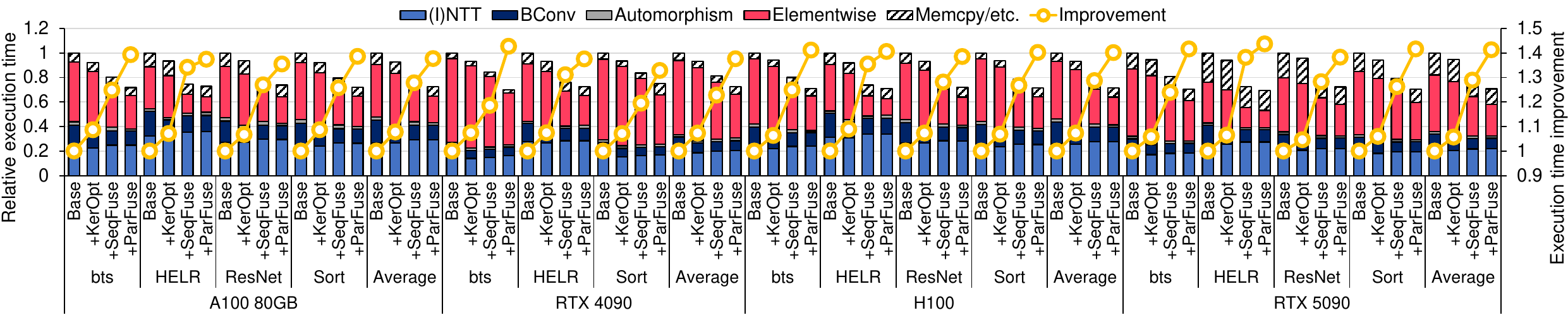}
     \vspace{-0.04in}
     \caption{Performance and execution time breakdown of \boot, HELR, ResNet, Sort, and their average.
     The incremental optimization configurations (Base, +KerOpt, +SeqFuse, and +ParFuse) are defined in Table~\ref{tab:ablation_config}.}
     \Description{}
     \vspace{-0.02in}
     \label{fig:ablation}
\end{figure*}

For the 48/24-limb case, \NAME outperforms the fastest open-source library, Phantom~\cite{tdsc-2024-phantom}, achieving 1.71$\times$ faster HMult, 1.76$\times$ faster HRot, and 1.64$\times$ faster rescaling.
Similarly, for the 24/12-limb case, \NAME achieves 1.73$\times$, 1.76$\times$, and 1.53$\times$ improvements, respectively.
As the benefits of our 25-30 prime system are excluded, these gains primarily stem from 32-bit execution and the kernel optimizations (\S\ref{sec:kernel}) that reduce the computational overhead.
While our kernel fusion methods also contribute, their impact is limited for these basic mechanisms with straightforward execution sequences.
For \hadd, which consists solely of element-wise operations, performance is constrained by memory bandwidth, resulting in minimal variation across the libraries.

\setlength{\tabcolsep}{2pt}
\begin{table}[t]
\centering
\small
\caption{Incremental optimization configurations used in ablation study (Fig.~\ref{fig:ablation}).
SMR and RR denote signed Montgomery reduction and rational rescaling, while EOT and LR represent extended on-the-fly twiddle factor generation and lazy reduction. \seqf{X} and \parf{X} represent sequential and parallel fusion optimizations detailed in \S\ref{sec:seq}.}
\vspace{-0.07in}
\label{tab:ablation_config}
     \begin{tabularx}{0.7\columnwidth}{>{\centering\arraybackslash}m{0.2\columnwidth}|>{\centering\arraybackslash}X}
    \toprule
    Config. &  Optimizations\\
    \midrule
    Base & SMR, RR, \seqf{1}, \parf{1--3} \\
    +KerOpt & +EOT, +LR\\
    +SeqFuse & +\seqf{2,3} \\
    +ParFuse & +\parf{4,5} \\
    \bottomrule
    \end{tabularx}
\vspace{-0.02in}
\end{table}

\subsection{Ablation Study}
\label{sec:eval:ablation}

We conducted an ablation study to evaluate the workload-level impact of our proposed kernel optimizations and fusion techniques.
Our baseline already incorporates several techniques from prior work, including signed Montgomery reduction (SMR), rational rescaling (RR), and specific fusion methods (\seqf{1}, \parf{1--3}).

We evaluated the workload performance of \NAME while incrementally applying three types of optimizations.
First, we applied kernel optimizations (+KerOpt), which include the extended on-the-fly twiddle factor generation for NTT and lazy reduction for the BConv kernel.
Second, we added sequential fusion (+SeqFuse), which includes the fusion for Montgomery conversion (\seqf{2}) and rescaling (or ModDown) (\seqf{3}) in \S\ref{sec:seq:sequential}.
Finally, we incorporated parallel fusion (+ParFuse), which includes the fusion for automorphism (\parf{4}) and linear transforms (\parf{5}) in \S\ref{sec:seq:parallel}.
These optimization configurations are summarized in Table~\ref{tab:ablation_config}.

Fig.~\ref{fig:ablation} shows the performance impact of our kernel optimizations and fusion techniques.
Kernel optimizations alone provide modest improvements of 5--7\% in average workload performance. 
These improvements are limited because our kernel optimizations primarily target NTT and BConv operations, while element-wise operations constitute the dominant portion of execution time at the workload level.
Element-wise operations account for 53\% of \boot time on an H100 and even 68\% on an RTX 4090, which features lower memory bandwidth and higher computational throughput.

The dominance of element-wise operations makes our fusion techniques particularly effective.
As sequential fusion reduces the memory accesses for basic mechanisms, including rescaling, \hmult, and \hrot, it leads to additional 18--22\% average workload performance improvements across the four GPUs.
Furthermore, as \boot includes linear transforms, our parallel fusion optimizations tested here benefit all FHE workloads, leading to additional 12--23\% average workload performance improvements.

Overall, our kernel optimizations and fusion techniques achieve average workload speedups of 1.38--1.41$\times$ across the four GPUs.

\section{Related Work \& Discussion}
\label{sec:related}
Although abundant GPU studies attempted to accelerate basic CKKS mechanisms~\cite{tdsc-2024-phantom, tc-2022-carm, access-2021-demystify,ipdps-2022-intel-gpu-ckks, ipdps-2023-ckks-gpu-fp, iacr-2024-heongpu, github-liberate_FHE}, only few~\cite{tches-2021-100x, discc-2024-fhe-cnn, heaan-latest, micro-2023-gme, hpca-2023-tensorfhe, hpca-2025-warpdrive} fully implemented \boot to demonstrate their applicability to FHE CKKS.
100$\times$~\cite{tches-2021-100x} identified that element-wise operations become the bottleneck in \boot and developed basic kernel fusion techniques for them.
Park et al.~\cite{discc-2024-fhe-cnn} used the HEaaN-GPU library~\cite{heaan-latest} to implement CNN workloads.
GME~\cite{micro-2023-gme} developed a custom hardware solution for FHE on top of the AMD MI100 GPU architecture.

Although \NAME demonstrates superior performance compared to prior studies, we may adopt orthogonal optimizations from them for further acceleration.
For example, we may utilize the tensor cores in recent NVIDIA GPU architectures.
TensorFHE~\cite{hpca-2023-tensorfhe} utilized the 8-bit integer datapath within the tensor cores to emulate 32-bit integer operations for (I)NTT.
WarpDrive~\cite{hpca-2025-warpdrive} improved upon TensorFHE by decomposing the (I)NTT matrix in a more fine-grained manner and utilizing both the int32 cores and the tensor cores.

Other studies~\cite{iiswc-2020-ntt,seed-2022-poly-gpu, pact-2021-ntt-tensor,tjs-2021-gpu-ntt,tjs-2021-gpu-ntt-nussbaumer} also optimized (I)NTT on GPUs.
Kim et al.~\cite{iiswc-2020-ntt} addressed the memory bandwidth bottleneck in (I)NTT through OT (\S\ref{sec:kernel:specific}).
Shivdika et al.~\cite{seed-2022-poly-gpu} proposed a modified modular reduction adequate for GPUs.
Goey et al.~\cite{tjs-2021-gpu-ntt} utilized the warp shuffle instructions to exchange the twiddle factors between threads.
Durrani et al.~\cite{pact-2021-ntt-tensor} used warp shuffle instead for exchanging polynomial data elements, addressing bank conflict issues in shared memory.
Meanwhile, Lee et al.~\cite{tjs-2021-gpu-ntt-nussbaumer} implemented the Nussbaumer algorithm on GPUs and compared it with (I)NTT.

Prior studies also accelerated CKKS with custom hardware designs using FPGA~\cite{hpca-2019-roy, asplos-2020-heax, hpca-2023-fab, hpca-2023-poseidon, hpca-2025-effact, hpca-2023-fxhenn} or ASIC~\cite{micro-2021-f1, isca-2022-bts, isca-2022-craterlake, micro-2022-ark, isca-2023-sharp, hpca-2025-anaheim}.
Apart from ASIC designs, which still require more research and development for realization, FPGA has recently become a popular platform due to its high flexibility and programmability.
FAB~\cite{hpca-2023-fab}, Poseidon~\cite{hpca-2023-poseidon}, and EFFACT~\cite{hpca-2025-effact} are notable FPGA implementations that support FHE CKKS with \boot.
However, due to the low hardware operating frequency and resource limitations, they reported modest performance values compared to \NAME.

\section{Conclusion}
\label{sec:conclusion} 

To resolve the high computational and memory overhead of FHE, we have developed \NAME, a swift GPU library for end-to-end execution of FHE workloads.
\NAME features a novel systemized RNS construction method named 25-30 prime system, architecture-aware 32-bit kernel optimizations, and optimized operational sequence implementations based on extensive kernel fusion;
these features are provided as a full-fledged GPU library with a high-level interface.
\NAME shows 2.18--4.45$\times$ faster workload execution times compared to WarpDrive~\cite{hpca-2025-warpdrive}, outperforms custom FPGA designs~\cite{hpca-2025-effact, hpca-2023-fab, hpca-2023-poseidon}, and achieves 1.53--1.76$\times$ speedups over Phantom~\cite{tdsc-2024-phantom} for basic mechanisms.

\section*{Acknowledgments}
We appreciate the invaluable feedback from Sangpyo Kim, Jaiyoung Park, Hyesung Ji, and Hyunah Yu.
This research was in part supported by Institute of Information \& communications Technology Planning \& Evaluation (IITP) grant funded by the Korea government (MSIT) [RS-2021-II211343, RS-2023-00256081, RS-2025-02217656].
Jongmin Kim is with the Interdisciplinary Program in Artificial Intelligence (IPAI), Seoul National University (SNU).
Wonseok Choi is with the Department of Intelligence and Information (DII), SNU.
Jung Ho Ahn, the corresponding author, is with DII and IPAI, SNU.

\balance

\bibliographystyle{ACM-Reference-Format}
\bibliography{refs}

\end{document}